\documentstyle[12pt]{article}
\def\np{Nucl.\ Phys.\ }

\def\prl{Phys.\ Rev.\ Lett.\ }

\def\prb{Phys.\ Rev.\ B }
\begin{document}
\setlength{\baselineskip}{.375in}
\vspace{1in}
\begin{center}
\Large\bf
Theory of the half-filled Landau level\\
\vspace{0.75in}
%\large\rm N. R{\normalsize EAD}\\
\large\sc N. Read\\
\vspace{0.325in}
\normalsize\em
Departments of Physics and Applied Physics, P.O. Box 208284,\\
Yale University, New Haven, CT 06520\\
\vspace{0.75in}
%\sc
%abstract
\end{center}
%\vspace{0.1in}
\rm
A recent theory of a compressible Fermi-liquid like state at Landau level
filling factors $\nu=1/q$ or $1-1/q$, $q$ even, is reviewed, with emphasis on
the basic physical concepts.
\begin{verbatim}
PACS Nos. 73.40 Hm, 03.70+k
February, 1994
\end{verbatim}
\newpage
The physics of interacting electrons in 2-dimensions in very high magnetic
fields has proved to be a rich subject since the discovery of the FQHE in 1982
\cite{tsui}. The initial motivation for study in this area was the expectation
that a Wigner crystal should form when the (areal) density of electrons is low
and the magnetic field is high, not only because, as at low magnetic field, the
Coulomb energy of repulsion dominates the kinetic energy at low density, but
also because if the magnetic field is high enough so that mixing of excitations
to higher Landau levels can be neglected then the kinetic energy itself becomes
essentially constant. In this limit the electrons would behave classically,
with their dynamics given by the ${\bf E}\times{\bf B}$ drift of their guiding
centre coordinates, {\it i.e.} drift along the equipotentials, the potential
being here given by the Coulomb potential of the other electrons. The Wigner
crystal is presumably the unique lowest energy, stationary state at a given
density in this classical problem. The surprise was that in fact, at accessible
densities, the quantum fluctuations cause this crystal to melt (as parameters
are varied) into some quantum fluid, and furthermore the nature of this fluid
depends on the commensuration between the density and the magnetic field, in an
essentially quantum mechanical way. In terms of the Landau level filling factor
$\nu\equiv n\Phi_0/B$ (where $n$ is the density, $B$ is the magnetic field
strength and $\Phi_0=hc/e$ is the flux quantum), this was manifested by
quantum Hall effect (QHE) plateaus at $\nu=p/q$ a rational number with an odd
denominator $q$ ($p$ and $q$ have no common factors) in all except a handful
of special cases observed more recently.

A theory that explained many features of this picture, including the ``odd
denominator rule'' was quickly forthcoming \cite{laugh,haldhier,halphier}.
Laughlin's
states at $\nu=1/q$, $q$ odd, and the hierarchical extension to all $\nu=p/q$,
$q$ odd, are incompressible fluids capable of exhibiting a QHE plateau at
$\sigma_{xy}=(p/q)e^2/h$. They also possess novel elementary excitations called
quasiparticles, which carry a fraction $\pm 1/q$ of the charge on an electron,
and have fractional statistics, that is the phase of the wavefunction changes
by $e^{i\theta}$, $\theta/\pi=\pm p'/q$ when two quasiparticles are exchanged
adiabatically \cite{halphier,arovas,readhier,blokwen}. A finite energy
$\Delta$ is
required to create one of these excitations. The fractional QHE associated
with a particular state will be destroyed if the energy scale characterizing
the rms disorder exceeds about $\Delta$.

Some questions remained unanswered, however. While the hierarchy theory
predicted that states at $\nu$ with larger denominators would have smaller
gaps, it would seem, at least naively, that all fractions with the same
denominator would be equally strong, though it is possible that those near a
stronger, smaller denominator state might be overwhelmed, for example $3/11$
near $1/3$. As samples improved, it could be seen that in fact most fractions
observed were of the restricted form $\nu=p/(mp+1)$, $p\neq0$, $m\geq0$ even,
$=2$, $4$. It was not clear if the hierarchy could explain this, although it
could be simply a quantitative fact, without deeper explanation.
Furthermore, in the region near $\nu=1/2$, where $m=2$, $p$ is large in the
above formula, no fractional plateaus were seen, but there was a shallow
minimum in $\rho_{xx}$ that remained $\neq0$ as $T\rightarrow0$. A possibly
more profound theoretical question was, what is the nature of the ground state
at fillings not of the hierarchy form $p/q$, $q$ odd? Indeed, the construction
of states at $\nu=p/q$, $q$ odd that exhibit FQHE does not prove that FQHE
cannot occur at even $q$. The whole question of what {\em does} occur received
almost no attention in published work until the last few years (a notable
exception is the FQHE at $\nu=5/2$ and the Haldane-Rezayi proposal for its
explanation \cite{will1,haldrez}). One possibility is that there is no well
defined state at these fillings in the thermodynamic limit, that phase
separation into
domains of two nearby, stable FQHE states occurs instead. Another possibility
is a well-defined pure phase
constructed by taking a FQHE state at a nearby filling, and adding a low
density of, say, quasiholes which at low density could form a Wigner crystal.
Such a state could exist (with varying lattice constant) over a range of
densities. These possibilities may actually occur near some particularly stable
(large $\Delta$) fractions such as $1/3$, but seem less likely near $\nu=1/2$
since neighbouring states would have very small energy gaps. If
liquid states at non-FQHE filling factors exist, then they lay
outside the then-current theoretical understanding.

With these motivating remarks, I now abandon the historical approach in
order to develop the overall theoretical picture \cite{HLR,jain,rezread}
as logically as possible. I
will describe two alternative approaches. The first begins close to Laughlin's
original ideas, and contains explicit trial wavefunctions, as well as other
notions of the last few years. The other approach is field theoretical and
eventually arrives at the same conclusions, but may seem less physical at
first.
However, it is much more appealing for explicit analytical calculation. I will
also explain the relation with Jain's work on the hierarchy states.

Let us turn, then, to the first of these approaches. We make the usual
assumption that, at $T=0$, interaction energies $\sim \nu^{1/2}e^2/
\epsilon\ell$ are weak compared with the Landau level splitting
$\hbar\omega_c$, and so the electrons should all be in the lowest
Landau level (LLL) with spins polarized when $\nu<1$. We work in the plane with
complex coordinate
$z_j=x_j+iy_j$ for the $j$th electron. In the symmetric gauge \cite{laugh},
single particle wavefunctions in the LLL are $z^m \exp\{-\frac{1}{4}|z|^2\}$,
where we set $\hbar=1$, and the magnetic length $\ell=\sqrt{\hbar c/eB}=1$. $m$
is the angular momentum of the state. An $N$ particle wavefunction has the form
\begin{equation}
\Psi(z_1,\ldots,z_N,\bar{z}_1,\ldots,\bar{z}_N)=f(z_1,\ldots,z_N)e^{-\frac{1}{4}
\sum_i|z_i|^2}
\label{LLLstates}
\end{equation}
where $f$ is complex analytic and totally antisymmetric. If $f$ is homogeneous
of degree $M$ in each $z_i$, it has definite total angular momentum and
describes a (not necessarily uniform) droplet of radius $\sim\sqrt{2M}$. As a
function of each $z_i$, $f$ has $M$ zeroes which is also the number of flux
quanta $N_\phi$ enclosed by the circle of radius $\sqrt{2M}$. (Similarly, on a
closed surface such as the sphere, the number of zeroes of the LLL single
particle wavefunctions equals the number of flux quanta through the surface.)
Thus the filling factor $\nu=N/N_\phi$ $=$ the number of particles per flux
quantum.

Now if $f$ contains a factor
\begin{equation}
U(z)=\prod_i(z_i-z)
\label{U(z)}
\end{equation}
then {\em one} zero for each particle is located at $z$. ($U(z)$ can be viewed
as an operator, Laughlin's quasihole operator, acting on the remainder of the
wavefunction by multiplication to produce $f$.) This means that there are no
particles in the immediate vicinity of $z$, there is a depletion of charge
there. We describe this as a {\em vortex} at $z$, since as in a vortex in a
superfluid, the phase of the wavefunction winds by $2\pi$ as any $z_i$ makes a
circuit around $z$.

Given a state, we can obtain a valid new state by multiplying by $U(z)$. (This
increases $M$ by $1$, if $z=0$, but otherwise by an indefinite amount. But it
always increases $N_\phi$ by $1$, if $N_\phi$ is defined as the flux through
the region occupied by the fluid.) For reasonably uniform fluid states (such as
Laughlin's, or the Fermi liquid below) the vortex can be considered as an
excitation of the fluid. Now add in addition an electron. Clearly it is
attracted to the centre of the vortex, due to the density deficiency there.
Since there is no kinetic energy, it can certainly form a bound state.
Similarly it can bind to multiple vortices $U(z)^q$.

It is natural to consider the possibility that the ground state itself contains
electrons bound to vortices, since this will give a low energy. As each vortex
is added, $N_\phi$ increases by $1$. If we add $1$ electron for each $q$
vortices, we can form identical bound states, and if all zeroes are introduced
in this way, we will have $N_\phi=q(N-1)$, which implies $\nu=N/N_\phi
\rightarrow1/q$ as $N\rightarrow\infty$. This case is by far the simplest
to understand. In this state, there will be $q$ -fold zeroes as one electron
approaches another. It has long been realized that Laughlin's Jastrow-like
ansatz
\begin{equation}
f_{\rm L}=\prod_{i<j}(z_i-z_j)^q
\end{equation}
can be called a ``binding of zeroes to particles''.

However, the astute reader will already be aware that $f_{\rm L}$ is
antisymmetric only for $q$ odd. To solve this problem, and obtain a deeper
insight into Laughlin's state, we must examine the nature of the bound objects
more closely. Since we know the properties of an electron, we turn to the
vortices. First, the $q$-fold vortex carries a charge $-\nu q$ in a fluid state
at arbitrary filling factor $\nu$ (this includes our compressible Fermi liquid
state below, as well as the usual hierarchy states). This can be established by
Laughlin's plasma analogy \cite{laugh} or indirectly through adiabatic motion
of the vortex \cite{arovas}; the latter will be more useful here. A crude
version of this argument is that when a vortex is moved around adiabatically in
some given fluid state, the wavefunction picks up a phase, since the
definition (\ref{U(z)}) of $U(z)$ shows that it changes phase by $2\pi$ for
each particle about which it makes a circuit. Then if it makes a circuit
around a (nonselfintersecting) closed loop enclosing a region $D$ of area $A$,
it will pick up a phase $2\pi\int_D d^2z\,\rho(z)$ which reduces to $\nu A$ if
the density $\rho(z)$ is uniform $\rho=\nu/2\pi$. This is then identified as
the same result as would be obtained for a particle of charge $-\nu$ in the
magnetic field seen by the electrons.
Similarly, a $q$-fold vortex has a charge deficiency of $\nu q$,
which for $\nu=1/q$ is equivalent to a real hole, so the electron--$q$-vortex
composite at this filling has net charge zero, and behaves like a particle in
{\em zero} magnetic field.
But note that the vortex is actually
sensitive to the density of electrons, which can vary in space and time, even
when the external magnetic field and the average filling factor are fixed.

Now for the famous fractional statistics of vortices. If a vortex makes a
circuit around a loop enclosing another vortex, with the density otherwise
uniform, the missing charge around the vortex core will make a difference of
$2\pi \nu$ to the phase (independent of the size and shape of the loop, as long
as the vortices remain far enough apart). But a circuit is equivalent to two
exchanges, up to translations. So a similar calculation gives that adiabatic
exchange of two vortices produces a phase $\theta=\pi\nu$. For two $q$-fold
vortices, the result is $\pi q^2\nu$ which at $\nu=1/q$ is $\pi q$. This shows
that at these fillings, $q$-fold vortices are fermions for $q$ odd, and bosons
for $q$ even. Hence the electron--$q$-vortex bound state is a boson for $q$
odd, and a fermion for $q$ even. For $q$ odd, we can now argue
that the bosons in zero magnetic field can Bose condense (at $T=0$) into the
zero-momentum state and that this
is the interpretation of the Laughlin state \cite{read}. If $\psi^\dagger$
creates an
electron in the LLL, then $\psi^\dagger(z)U^q(z)$ creates a boson, and the
condensate is obtained by letting $\int d^2z\, \psi^\dagger(z)U^q(z)
\exp\{-\frac{1}{4}|z|^2\}$ act
repeatedly on the vacuum. This produces exactly the Laughlin state \cite{read}.
Note that condensation arises because particles try to minimise their kinetic
energy $\sim k^2$. We need to show that the bound objects do in fact have such
an {\em effective} ``kinetic'' energy. The true kinetic energy of the electrons
has been quenched, so this term in the effective Hamiltonian for the bound
objects can only arise from electron-electron interactions. Before discussing
how this arises, let us pursue the consequences. At $\nu=1/q$, $q$ even, the
bound objects are fermions in zero net field, which cannot Bose condense, but
can form a Fermi sea. This is then the proposal for a compressible state at
these fillings. It will be compressible because it can be shown that
incompressibility, and the quantized Hall effect, in the case of bosonic bound
states is a consequence of Bose condensation (superfluidity), which does not
occur in a Fermi sea unless BCS pairing occurs, in which case the state
becomes a QHE state. The Haldane-Rezayi and Pfaffian states arise in this
way \cite{mooreread}.

I now return to the dynamics of the bound objects, or ``quasiparticles''. We
have argued that there is an attraction between an electron and a $q$-fold
vortex. In a uniform background, this will take the form of a central
potential. The minimum is at the centre, meaning the electron is exactly at the
zeroes of the many-particle wavefunction. Now at filling factor $1/q$, we need
to consider quasiparticles in plane wave states with wavevector $\bf k$. These
can be created by acting on a fluid with $\int d^2z\, e^{i{\bf k}\cdot{\bf r}}
\psi^\dagger(z)U^q(z)\exp\{-\frac{1}{4}|z|^2\}$. In the wavefunction,
this means a factor $e^{i{\bf k.r}_j}$ in the term where the $j$th electron is
the bound in the state with wavevector $\bf k$. Now it turns out that this
factor, acting by multiplication on some given state, displaces
the $j$th particle by $ik$, where $k=k_x+ik_y$ (recall that the
magnetic length $\ell$ is 1). This arises because in the Hilbert space of many
particle states of the form (\ref{LLLstates}), $\bar{z}_j$ acts on $f$ as
$2\partial/\partial z_j$ which generates displacements \cite{girvinjach}. A
quasiparticle with wavevector zero would have the electron exactly at the
zeroes of the wavefunction. So a quasiparticle with wavevector $\bf k$ has the
electron displaced by $|k|$ from the centre of the vortex. The electron and
vortex experience a
potential $V(|k|)$ due to the Coulomb repulsion of the electron by the other
electrons, which are excluded from the vortex core. A good understanding is now
achieved semiclassically. The electron will drift along an equipotential of
$V(|k|)$. From the preceding discussion, at $\nu=1/q$ the $q$-fold vortex
experiences a magnetic field of the same strength as the electron, and so it
will also drift with the same speed but in the opposite sense relative to the
gradient of the potential, due to the opposite sign of its effective charge.
This means that both components of the pair drift in the same direction,
perpendicular to the vector connecting their centres, so that their separation
($=|k|$) remains  constant. The picture is like that of oppositely charged
particles in  a  magnetic field, which can drift in a straight line as a pair.
The energy of our pair is $V(|k|)$ and the velocity is $\propto \partial
V/\partial |k|$ as it should be. Near the bottom of the potential, it will be
quadratic, and we can obtain an effective mass $\sim (\partial^2
V/\partial |k|^2)^{-1}$ due to the interactions (a similar calculation was
performed in \cite{read}). This shows clearly that the
interactions favour condensation of the quasiparticles to minimize this
effective kinetic energy. In the $q$ even case, the quasiparticles are fermions
and must have distinct $\bf k$'s, filling a Fermi sea. Then in
the ground state not all the zeroes of the wavefunction are precisely on
the electrons but some are
displaced by amounts up to $k_F$, determined in the usual way by the density,
$k_F^2/4\pi=1/2\pi q$. For $q$ large these displacements $\sim q^{-1/2}$ are
small compared with the interparticle spacing $\sim q^{1/2}$. Thus a trial
state of this form may not be much worse energetically than the Laughlin state
at nearby odd $q$. This, I believe, is then the essential reason why this idea
has a good chance of being the correct many-body ground state: it is due to the
good correlations that produce a low Coulomb energy. Of course, if $q$ is very
small, this picture might break down, but in fact we have good reason to
believe it holds for $q$ as small as $2$. This argument also tells us that
low-lying excited states are obtained by increasing the wavevector of a
quasiparticle in the Fermi sea. For $q$ large, the Fermi velocity will be
determined by the same effective mass as near the bottom of the sea; otherwise
we must take the derivative at $k_F$.

To consider collective effects it is necessary to go beyond an independent
quasiparticle picture. There are important long range interactions that can be
described as gauge fields. We will examine these in the context of the field
theoretic approach later. As a test of the above ideas, we can perform
numerical diagonalization of small systems at, say, $\nu=1/2$. This has been
done recently in a paper by E. Rezayi and the author \cite{rezread}. Excellent
agreement of trial states suggested by the above ideas is found with the
exact wavefunctions of low-lying states. The trial wavefunctions were
generated on the sphere, but on the plane would be roughly
\begin{equation}
\Psi={\cal P}_{\rm LLL}\det M\,\prod_{i<j}(z_i-z_j)^q
e^{-\frac{1}{4}\sum|z_i|^2}.
\label{trialstate}
\end{equation}
The matrix $M$ has elements that are essentially plane waves $M_{ij}\sim
e^{i{\bf k}_i\cdot{\bf r}_j}$ for the
quasiparticles, filling the Fermi sea. ${\cal P}_{\rm LLL}$ projects all
electrons to the LLL. As explained above, the plane wave factors then act as
operators within the LLL, on the Jastrow factor $\prod(z_i-z_j)^q$ which if not
modified would be the Laughin state for bosons at $\nu=1/q$. The Slater
determinant makes the state totally antisymmetric.
The simple product form is similar to Jain's states off half-filling
\cite{jain}; it
differs a little, but inessentially, from the present author's first idea
(in 1987) of building
the Fermi sea by acting with Fourier components of $\psi^\dagger(z)
U^q(z)\exp\{-\frac{1}{4}|z|^2\}$, which was suggested by the analogy with the
Laughlin states at $q$ odd \cite{read}.
The numerical calculations also showed that the two-particle correlation
function $g(r)$ in the ground state possesses ({\it i\/}) a large
``correlation hole'' at short distances $r$, consistent with the argument
that zeroes are bound close to the electrons, and ({\it ii\/})
oscillations at large $r$, perhaps of the form $r^{-\alpha}\sin 2k_Fr$
asymptotically, as in a two-dimensional Fermi gas, which has $\alpha=3$.

We now turn to the field theoretic approach. It begins with the observation
that in 2 dimensions, particles of any (fractional) statistics can be
represented by charged particles of other statistics attached to
$\delta$-function flux tubes of a certain size \cite{wilczek}. A charged
particle dragged adiabatically around a flux tube (or {\it vice versa})
picks up an Aharonov-Bohm
phase of $2\pi$ times the product of charge of the particle and the number of
flux quanta in the
tube. Thus if identical particles are attached to identical flux tubes, a
circuit of one composite around the other produces a phase $4\pi$ times the
charge times the flux, since each particle sees the other flux. For an
exchange, we get only half this. In addition, the exchange of identical
particles produces a phase $e^{i\theta}$ due to the statistics of the particles
themselves. Thus fermions or bosons attached to fractional flux tubes can be
used to model anyons, where the total phase obtained in an exchange is
fractional \cite{wilczek}. In the fractional quantum Hall effect we change our
terminology slightly because we view flux tubes as operators that act only on
particles already present. Thus a composite is introduced by adding first a
flux
tube, then a particle is added at the same point. The phase produced by an
exchange is then only $\pi$ times the charge times the flux in each composite,
plus the phase due to exchanging the particles themselves. Thus, for example,
we can say that attaching two flux quanta to each boson in a system leaves the
composites still as bosons, or doing the same to fermions leaves them fermions.
Then in the high field situation of interest here, we may represent electrons
as $q$ (integer) flux tubes attached to some other particles. The latter must
be bosons if $q$ is odd, and fermions if $q$ is even, in order to reproduce the
fermi statistics of the electrons. The flux tubes can be represented by a
``fictitious'' vector potential $\bf a$, not to be confused with the physical
vector poetential $\bf A$ representing the constant external field, which obeys
\begin{equation}
\nabla\times {\bf a}=2\pi q \rho
\end{equation}
and the density $\rho({\bf r})=\sum \delta^{(2)}({\bf r}_i-{\bf r})$.

The next step, used in several similar problems \cite{zhang,laughany,jain,HLR},
is a mean field approximation. The
new fermions (for $q$ even) see the constant background magnetic field, and
the $q$ flux tubes attached to the other fermions. If the quantum mechanical
state has  a uniform density (in the quantum average), the latter becomes a
constant field whose sign we can choose to be opposite to the external field.
In particular, if $\nu=1/q$, the fields cancel exactly. The fermions now see no
net field, so they can form a Fermi sea, which does have a uniform density, as
we assumed. We have arrived at the physical picture of the compressible Fermi
liquid-like state, where we can say loosely that the fermions are electrons
plus $q$ flux quanta. But notice that this should not be taken too literally,
since we have actually just made a transformation. The real flux due to the
external field remains uniform, not bunched up into flux tubes attached to the
electrons. It is really vorticity that is bound to the electrons, as discussed
above. In mean field theory the wavefunction for the electrons is simply
\begin{equation}
\Psi_{\rm MF}=\det M\,\prod_{i<j}(z_i-z_j)^q/|z_i-z_j|^q
\label{MFstate}
\end{equation}
with $M$ as before and no projection. This function is {\em not} all in the
LLL. We can see that the effect of the transformation and the mean field
approximation is to build the right kind of phase factors into the
wavefunction, the same as possessed by $q$ vortices on each electron.
The factor $\prod |z_i-z_j|^q$ needed to recover eq.\ (\ref{trialstate}) can in
fact be obtained from fluctuations about mean field, at least in a
long-wavelength sense \cite{zhangrev}.

Excited states again involve creation of particle-hole pairs.
In mean field approximation, the effective mass of fermion excitations near
the Fermi surface is simply the bare electron mass $m$, since that is what
appears in the Hamiltonian (we do not attempt to impose the LLL constraint).
Consideration of collective oscillations of the system leads to the correct
cyclotron frequency $\omega_c=eB/mc$ which according to Kohn's theorem cannot
be renormalized. However, under the conditions stated at the beginning of this
article, the low-energy fermi excitations should have an effective mass $m^*$
determined by inter-electron interactions only. Part of the resolution of this
problem is that if fluctuations renormalize the effective mass, as they usually
do in Fermi liquids, then there will also be a Landau interaction paramater
$F_1$ that obeys a relation $m^{-1}=m^{*-1}+F_1$ which guarantees that the
collective mode frequency is unrenormalized (like the plasma frequency which is
unrenormalized in the usual electron gas at zero magnetic field) \cite{HLR}.
A more serious problem is that interactions seem to play no role in the mean
field theory. The prediction would be the same, even for noninteracting
electrons. The Fermi liquid mean field state has finite compressibility, due to
the mass $m$, even though in this case a partially filled Landau level should
have infinite compressibility! (The same holds when this approach is applied to
the FQHE states for $q$ odd, where the compressibility vanishes.) These
problems
can be resolved by understanding in what limit the approach is valid. Mean
field theory is good, and fluctuations in the effective magnetic field
$\nabla\times({\bf A}+{\bf a})$ are small, when $q$ is small. But for us
$q$ is an even integer $\geq2$. However if we replace the original problem of
electrons in a magnetic field by that of anyons in a magnetic field, then we
can make $q$ small.
We choose to study anyons with statistics $\theta=\pi(1+q)$ (mod
$2\pi$) and filling factor $1/q$, for $q>0$.
As $q\rightarrow 0$, the magnetic field goes to zero, but for all $q$ we can
still map the problem of anyons to one of fermions in zero net average
magnetic field. So at $q=0$ we reach fermions at zero magnetic
field and we attach zero flux to convert them to fermions! At this point there
can be no fluctuations in the effective magnetic field, since there is no
attached flux. As we increase the external
field, we must attach more flux to map to fermions in zero field, so the
statistics must change accordingly. When $\theta$ has made a full circle from
$\pi$ back to $\pi$, we recover fermions (electrons) but now at a high field
and filling factor $1/q$, $q=2$. We can continue and reach other even
denominator states for electrons.
This idea is an adaptation of that in
\cite{greiterwilzcek}, except that they use it to argue that certain states
exist by adiabatic continuation, while I use it only for a perturbation
expansion in the fluctations, {\it i.e.} in powers of $q$. Notice that in the
limit $q=0$, the system is a Fermi liquid, and the effective mass may be
close to the bare mass if interactions are weak compared to the Fermi energy.
A similar statement applies to the anyons for $q$ small.
It is only at $q=2$ that we can argue that the compressibility must be
infinite when interactions are set to zero, because only for fermions (such as
the electrons) or charged bosons, {\em and not for anyons},
can we argue that the many-particle states of a noninteracting system are
(anti-)symmetrized products of Landau level states.

After that rather technical paragraph, we return to simpler discussion. If the
filling factor for electrons deviates from $\nu=1/q$, $q$ even, then in either
approach above the quasiparticles will see a nonzero net magnetic field
$B_{\rm eff}=B-B_{1/2}=\nabla\times({\bf A+a})$. If the
quasiparticles fill an integer number $p$ of Landau levels due to this
effective field, there will be a gap in the excitation spectrum (at least in
mean field theory, but in fact it survives fluctuations) and we will obtain an
integer QHE for the fermions, which is an FQHE for the electrons, with
$\sigma=(e^2/h)p/(qp+1)$. This of course is Jain's picture \cite{jain} of the
``main sequence'' of fractional QHE states as observed in experiments. Jain's
approach was a hybrid of the approaches above. He begins with the
transformation involving $\delta$-function fluxes, but then simply modifies the
mean field wavefunctions (\ref{MFstate}) into the form (\ref{trialstate}) with
$M_{ij}=$ Landau level wavefunctions. The argument of binding zeroes to
particles to obtain a low Coulomb energy is not used. We believe the
$\nu=1/q$ case with exactly $q$ vortices per particle gives valuable insight
into the reasons for the stability of the states, even away from this filling.
Also the notion of the effective mass emerges clearly in the Fermi liquid
state. We predicted \cite{HLR} that energy gaps in FQHE states near
half-filling should scale as $1/p$ as $p\rightarrow\infty$, since the gap can
be interpreted as an effective cyclotron energy $\sim B_{\rm eff}/m^*$
once we recognize the existence of the Fermi liquid, with a well-defined
effective mass that controls excitation energies, as the limiting behaviour.
This prediction has received experimental support \cite{du}.

The above arguments about dynamics of the quasiparticles at $\nu=1/2$ also
generalizes to this case. While the electron and the $q$-fold vortex still see
the same potential $V(|k|)$, the drift velocities are different because that
for the vortices is fixed by the electron density, not the true magnetic field.
The separation still remains constant, so like the back wheels of a car turning
a corner, the bound pair moves (semiclassically) on two concentric circles
separated by $|k|$, giving an effective cyclotron radius for the
quasiparticles. This radius is given by the usual formula $\Phi_0 k/2\pi B_{\rm
eff}$ for a particle of charge 1 in a magnetic field $B_{\rm eff}$.

The reader is cautioned that away from half-filling the quasiparticles which
seem to be fermions in the mean field approach in fact have their statistics
modified because of screening effects of the fluctuations. They also acquire
fractional charge, whereas at half filling there is perfect screening by the
response of the fluid to the $\delta$-function of flux on the fermion. This
effect is in fact described by the extra amplitude factors present in the
wavefunctions (\ref{trialstate}). The arguments given above
for electron--$q$-vortex bound states at $\nu=1/q$ extend to other fillings to
predict a net charge $1-q\nu$ and statistics $\theta=\pi(q^2\nu-1)$. These
properties of fermions excited to higher ``quasiparticle Landau levels'' at
Jain's fractions are identical to those of the quasiparticle
excitations in the hierarchy scheme, which is one of the main evidences for
the equivalence of these two approaches \cite{readhier,blokwen}.

Space allows only a brief further discussion. An outstanding series of
experiments has been performed by Willett and coworkers \cite{will2}. Their
technique, using
the propagation of surface acoustic waves across the surface of the device,
probes the longitudinal conductivity of the 2 dimensional electron gas at
finite
wavevectors and frequencies (this and other experiments \cite{kang} will be
discussed at this meeting by Dr. Stormer). Since the velocity of the waves is
slow compared with the Fermi velocity of our fluid, we can consider zero
frequency, finite wave vector. The theory, treating the response of the
fermionic quasiparticles in an RPA-like approximation, shows that
$\sigma_{xx}(k)$
increases linearly with wavevector $k$ in the compressible state, as observed.
This results from the {\em transverse} conductivity $\sim 1/k$ for a
conventional Fermi liquid, on including the long-range effects of the gauge
field $\bf a$. (Not only can $B_{\rm eff}=2\pi q(n-(2\pi q)^{-1}))$ fluctuate
in space and time, so can $E_{\rm eff}=2\pi q J$, the effective electric field
due to the current $J$ in the perpendicular direction.) It also predicted
resonances in the response when the wavevector
is an integer multiple of the inverse cyclotron radius for the semiclassical
motion of the quasiparticles close to but just off half-filling. This provides
a measure of the Fermi wavevector. The recent experimental observation of
these resonances confirms the (nontrivial!) existence of a Fermi surface for
the charge-carrying excitations and excellent agreement of $k_F$ with the
expected value ($\sqrt{2}$ times that in zero magnetic field, because of
spin) is obtained.

Open questions: a direct measure of the effective mass $m^*$ at half-filling
would be most welcome. Theoretically, a controversy remains about the
possible partial breakdown of Fermi liquid theory, including behaviour of
$m^*$, due to the fluctuations of $\bf a$ \cite{HLR}.

I am grateful to P.A. Lee, B.I. Halperin and E. Rezayi for collaborations. This
work was supported by NSF-DMR-9157484 and by the A.P. Sloan Foundation.

\vspace*{\fill}
\end{document}